\documentclass[aoas,MSNbibl,nameyear,dvips]{arximspdf}
\usepackage{float}
\usepackage{dcolumn}
\usepackage{multirow}
\usepackage{graphicx}

\doi{10.1214/14-AOAS750} 
\volume{8}
\issue{3}
\pubyear{2014}
\firstpage{1690}
\lastpage{1712}
\docsubty{FLA}

\makeatletter
\newcolumntype{d}[1]{D{.}{.}{#1}}

\newproclaim{assumption}{Assumption}
\newtheorem{teo}{Theorem}
\floatstyle{ruled}
\newfloat{algo}{thp}{lop}
\floatname{algo}{Algorithm}
\makeatother

\begin{document}
\begin{frontmatter}

\title{Automatic estimation of flux distributions of astrophysical
source populations}
\runtitle{Estimation of astrophysical source populations}

\begin{aug}
\author[A]{\fnms{Raymond~K.~W.}~\snm{Wong}\thanksref{A1}\ead[label=e1]{rkwwong@ucdavis.edu}},
\author[A]{\fnms{Paul}~\snm{Baines}\thanksref{A1}\ead[label=e2]{pdbaines@ucdavis.edu}},
\author[A]{\fnms{Alexander}~\snm{Aue}\thanksref{A1,T1}\ead[label=e3]{aaue@ucdavis.edu}},
\author[A]{\fnms{Thomas~C.~M.}~\snm{Lee}\corref{}\thanksref{A1,T2}\ead[label=e4]{tcmlee@ucdavis.edu}}
\and
\author[B]{\fnms{Vinay~L.}~\snm{Kashyap}\thanksref{A2,T3}\ead[label=e5]{kashyap@head.cfa.harvard.edu}}
\runauthor{R.~K.~W. Wong et al.}
\affiliation{University of California, Davis\thanksmark{A1} and\\
Harvard--Smithsonian Center for Astrophysics\thanksmark{A2}}
\address[A]{R.~K.~W. Wong\\
P. Baines\\
A. Aue\\
T.~C.~M. Lee\\
Department of Statistics\\
University of California, Davis\\
4118 Mathematical Sciences Building\\
One Shields Avenue\\
Davis, California 95616\\
USA\\
\printead{e1}\\
\phantom{E-mail: }\printead*{e2}\\
\phantom{E-mail: }\printead*{e3}\\
\phantom{E-mail: }\printead*{e4}}
\address[B]{V.~L. Kashyap\\
Harvard--Smithsonian Center\\
\quad for Astrophysics\\
60 Garden Street\\
Cambridge, Massachusetts 02138\\
USA\\
\printead{e5}}
\end{aug}
\thankstext{T1}{Supported in part by NSF Grants 1209226 and 1305858.}
\thankstext{T2}{Supported in part by NSF Grants 1007520, 1209226 and 1209232.}
\thankstext{T3}{Supported in part by NASA Contract NAS8-03060 to the
Chandra X-Ray Center.}

\received{\smonth{5} \syear{2013}}
\revised{\smonth{4} \syear{2014}}

%
\begin{abstract}
In astrophysics a common goal is to infer the flux distribution of
populations of scientifically interesting objects such as pulsars or
supernovae. In practice, inference for the flux distribution is often
conducted using the cumulative distribution of the number of sources
detected at a given sensitivity. The resulting ``$\log(N>S)$--$\log
(S)$'' relationship can be used to compare and evaluate theoretical
models for source populations and their evolution. Under restrictive
assumptions the relationship should be linear. In practice, however,
when simple theoretical models fail, it is common for astrophysicists
to use prespecified piecewise linear models. This paper proposes a
methodology for estimating both the number and locations of
``breakpoints'' in astrophysical source populations that extends beyond
existing work in this field. 

An important component of the proposed methodology is a new interwoven
EM algorithm that computes parameter estimates. It is shown that in
simple settings such estimates are asymptotically consistent despite
the complex nature of the parameter space. Through simulation studies
it is demonstrated that the proposed methodology is capable of
accurately detecting structural breaks in a variety of parameter
configurations. This paper concludes with an application of our
methodology to the \textit{Chandra} Deep Field North (CDFN) data set.
\end{abstract}

%
\end{frontmatter}

\setcounter{footnote}{3}

\section{Introduction}

The relationship between the number of sources and the threshold at
which they can be detected is an important tool in astrophysics for
describing and investigating the properties of various types of source
populations. Known as the $\log N$--$\log S$~relationship, the idea is
to use the number of sources $N(>S)$ that can be detected at a given
sensitivity level $S$, on the $\log$--$\log$ scale, to describe the
distribution of source fluxes. In simple settings and under restrictive
assumptions a linear relationship between the log-flux and the
log-survival function can be derived from first principles.
Traditionally, astrophysicists have therefore examined this
relationship by characterizing the slope of the log of the empirical
survival function as a function of the log-flux of the sources.

One of the first examples of the $\log N$--$\log S$ relationship
being derived from first principles is in~\citet{Schuer57}. It is
shown that if radio stars are uniformly distributed in space, then the
number with intensity exceeding a threshold $S$ is given by
$N(>S)\propto S^{-3/2}$. Importantly, the relationship holds
irrespective of several factors such as luminosity dispersion and the
reception pattern of the detector. The derived relationships therefore
allow for researchers to test for departures from specific theories.
For example,~\citet{Hewish61} uses the derived relationship to
infer a nonuniform distribution of sources for a particular population.

Other examples of~$\log N$--$\log S$~analyses include~\citet
{Guettaetal2005}, who use the relationship for Gamma Ray Bursts (GRBs)
to constrain the structure of GRB jets. By comparing the $\log
N$--$\log S$~relationship for observed data to the predicted $\log
N$--$\log S$~relationship under different physical models for GRB jets,
the authors are able to uncover limitations in the physical models. The
$\log N$--$\log S$~curves have also been used to constrain cosmological
parameters using cluster number counts in different passbands; see, for
example, \citet{MathiesenEvrard1998} and \citet
{Kitayamaetal1998}. Other applications of $\log N$--$\log S$ modeling
include the study of active galactic nuclei (AGNs). For
example, \citet{Mateosetal2008} use the $\log N$--$\log
S$ relationship over different X-ray bands to constrain the population
characteristics of hard X-ray sources.

Under independent sampling, the linear $\log N$--$\log S$ relationship
corresponds to a Pareto distribution for the source fluxes, known to
astrophysicists as a power-law model. Despite the unrealistic
assumptions in the derivation, the linear $\log N$--$\log
S$ relationship does have strong empirical support in a variety of
contexts, for example, \citet{KenterMurray03}. In addition to its
simplicity, the power-law model also retains a high degree of
interpretability, with the power-law exponent often of direct
scientific interest. As a result of this simplicity and
interpretability, the power-law model forms the basis of most $\log
N$--$\log S$ analyses despite its many practical limitations in the
ability to fit more complex data sets.

To address the limitations of this simple model, astrophysicists have
also experimented with a variety of broken power-law models. This is
particularly important for larger populations or populations of sources
spread over a wide energy range. \citet{Mateosetal2008}
illustrate this by using both a two- and three-piece broken power-law
model to capture the structure of the $\log N$--$\log S$ distribution
across a wide range of energies. The basic idea of broken power-law
models is to relax the assumption that the log survival function is a
linear function of the log flux, and to instead assume a piecewise
linear function. This adds additional challenges in estimating the
location of the breakpoint and quantifying the need for the breakpoint
model above the simpler single power-law model. While recognizing the
need to have more flexible models for $\log N$--$\log S$ analyses, most
of the work in this area does not provide a coherent means to selecting
the location and number of breakpoints.

Similarly to the single power-law model, the broken power-law model can
be derived from first principles as a mixture of truncated and
untruncated Pareto distributions. The direct physical plausibility of
the model is not as complete as for the single power-law model, but the
model parameters, in particular, the slopes of the $\log N$--$\log
S$ relationship, can be used to draw conclusions about competing
theories. The broken power law provides a useful approximation that can
be used to model mixtures of populations of sources, as well as more
general piecewise-linear populations. Indeed, the broken power law has
empirical support in a variety of contexts both in
astrophysics [\citet{Mateosetal2008,Kouzu13}] and
outside [\citet{Segura13}].

There are many alternative generalizations of the single power law in
addition to the broken power law considered in this paper. For
example, \citet{Ryde99} considers a smoothly broken power-law
model that avoids the nondifferentiability introduced by the strict
broken power-law model. Other alternatives include mixtures of
log-normal distributions and power laws with modified tail behavior. In
addition to parametric methods, the flux distribution can also be
modeled nonparametrically. For the types of applications we are
considering here, the main goal is parameter estimation and model
selection to distinguish between single and broken power-law models.
The scientific interpretability of a nonparametric model for the $\log
(N>S)$--$\log(S)$ relationship is more complicated than the parametric
alternative, and such approaches have gained less traction in the
astrophysics community in the context of $\log N$--$\log S$ analyses.
Therefore, while a more flexible nonparametric fit is perhaps
statistically preferable, it is not as amenable to downstream science
as in other contexts where the goal is prediction rather than estimation.

Among all generalizations, the strict broken power law remains the most
popular alternative. This popularity is a result of the
interpretability of the model and the ease of translation from
statistical results to scientific interpretability. Despite the
popularity of the broken power-law model in the $\log N$--$\log S$
literature, there is currently no widely applicable and statistically
rigorous method framework for fitting broken power-law models to the
$\log N$--$\log S$ relationship to astrophysical source populations.

In this paper we provide an automatic method for jointly inferring the
number and location of breakpoints and the parameters of interest for
the $\log N$--$\log S$ problem. Our method allows astrophysicists to
reliably infer both the number and the location of breakpoints in the
$\log N$--$\log S$ relationship in a statistically rigorous manner for
the first time. This simultaneous fitting introduces new computational
challenges, so our method utilizes a new extension of the EM algorithm,
known as the interwoven EM algorithm (IEM) [\citet
{Baines10,Baines12}]. The IEM algorithm provides efficient and stable
estimation of the model parameters across a wide range of parameter
settings for a fixed number of breakpoints. To determine the number of
breakpoints, we then use an additional model selection procedure that
employs the power posterior technique of \citet{Friel-Pettitt08}
to accurately compute the log-likelihood of the candidate models.

The remainder of the paper is organized as follows. In Section~\ref
{secmotivation} we introduce the necessary background and statistical
formulation of the $\log N$--$\log S$ model. Section~\ref
{secestimation} provides details of our estimation procedure for a
fixed number of breakpoints, with Section~\ref{secchoiceB} outlining
our model selection procedure to determine the number of breakpoints
required. The performance of our method in terms of both parameter
estimation and identification of the number of breakpoints is detailed
in Section~\ref{secSims}. An application to data from the \textit
{Chandra} Deep-Field North X-ray survey is provided in Section~\ref
{secCDFN}. Large-sample theory is developed in Section~\ref
{seclargesample} and concluding remarks are offered in Section~\ref
{secConc}. Last, technical details are given in an online supplement
[\citet{Wong-Baines-Aue14}].

\section{Background and problem specification}\label{secmotivation}

Let $\mathbf{S}=(S_{1},\ldots,S_{n})^{T}$ denote a vector of the fluxes
(in units of erg~s$^{-1}$~cm$^{-2}$) of each of a population of $n$
astrophysical sources. For example, we may be interested in the flux
distribution of a selection of $n$ X-ray pulsars located in a specified
region of sky at a specified distance. The basic building block of our
method is the power-law model:
%
\begin{equation}
\label{eqpowerlaw} N(>S) = \sum^n_{i=1}I_{\{S_i>S\}}
\simeq\alpha S^{-\beta}, \qquad S>\tau.
\end{equation}
This specifies that the unnormalized survival function $N(>S)$ is
approximately a power of the flux $S$. The power-law exponent, $\beta
$, is the parameter of primary interest and provides domain specific
knowledge about the source populations. The lower threshold $\tau$ can
either be fixed according to the desired sensitivity level or estimated
from the data. Equivalently, taking the logarithm of both sides, (\ref{eqpowerlaw})~assumes a linear relationship between $\log(N(>S))$ and
$\log(S)$:
%
\begin{equation}
\label{eqlogpowerlaw} \log \bigl(N(>S) \bigr) \simeq\log(\alpha) - \beta\log(S), \qquad
S> \tau.
\end{equation}
In a statistical context, the theoretical power-law assumption
corresponds to assuming that the source fluxes follow a Pareto distribution:
\[
S_i \stackrel{\mathrm{i.i.d.}} {\sim}\operatorname{Pareto}(\beta,\tau), \qquad i=1,\ldots,n.
\]
In practice, the linear $\log N$--$\log S$, or Pareto, assumption is
not sufficient to describe the $\log N$--$\log S$ relationship for many
real data sets. There are several ways to generalize (\ref
{eqpowerlaw}), the most popular among astrophysicists being the broken
power-law model as illustrated in \citet{Virgo04} and \citet
{XMMNewton07}. The starting point of the broken power law is to
replace (\ref{eqpowerlaw}) with a monotonically decreasing piecewise
linear approximation. In the case of a two-piece model we assume
%
\begin{equation}
\label{eqtwobreak} \log \bigl(N(>S) \bigr) = \cases{ \displaystyle\log(
\alpha_{1}) - \beta_{1}\log(S), &\quad$\tau _{1}
< S \leq\tau_{2}$,
\vspace*{3pt}\cr
\displaystyle\log(\alpha_{2}) -
\beta_{2}\log(S), &\quad$S > \tau_{2}$,}
\end{equation}
where $\beta_{1}$ and $\beta_{2}$ are parameters of interest. Note
that as a result of the continuity and normalization constraints on
$\tau_{1}, \tau_{2}, \alpha_{1}, \alpha_{2}, \beta_{1}$ and $\beta
_{2}$, there are a total of 4 free parameters in this expanded
two-piece model. Applications of the broken power-law model in the
astrophysics community typically use either fixed numbers and locations
of the breakpoint(s) or selection via ad hoc procedures [\citet
{Trudolyubov02}]. The contribution of this paper is the proposal of an
automatic procedure for selecting the number and estimating the
locations of the breakpoints jointly with the parameters of interest.\looseness=-1

%
\begin{figure}[b]

\includegraphics{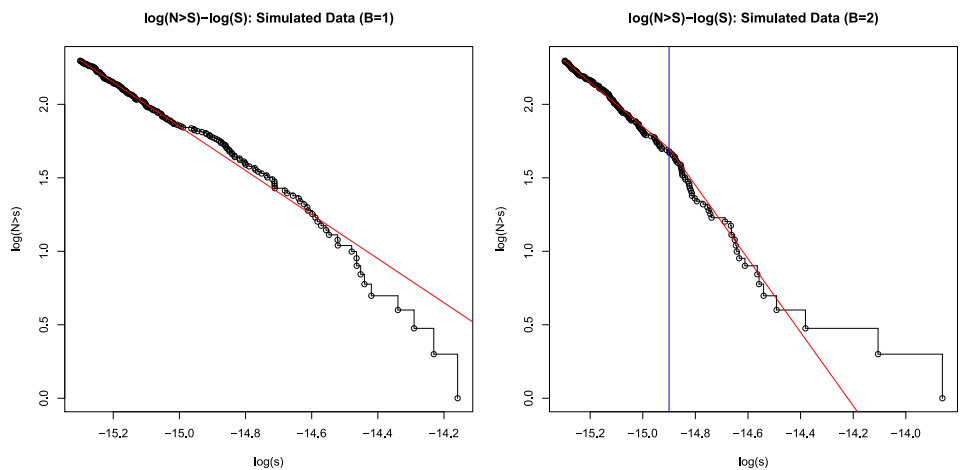}

\caption{Example simulations of flux distributions under the single
power-law model
\textup{(left)}, and the broken power-law model \textup{(right)}. In
practice, the
fluxes are not directly observed and must be inferred from count data
as described
in Section \protect\ref{secmotivation}. For the broken power-law
example, the vertical blue line corresponds to the location of the
breakpoint.}\label{figsimplots}
\end{figure}

Figure~\ref{figsimplots} depicts the $\log N$--$\log S$ relationship
for flux distributions simulated under a single power-law (left) and
broken power-law model (right). As may be expected, even under a
theoretically linear relationship, the empirical \mbox{$\log N$--$\log
S$-}curve regularly exhibits nonlinear features in the $\log N$--$\log
S$-space. Depending on the difference in the power-law slopes, the
breakpoint may be clearly visible or indistinguishable by eye. In
either case, it should be noted that much larger variations in the
$\log N$--$\log S$ relationship are to be expected in the lower right
part of the curves as a result of the $\log$--$\log$ scaling. As will be
seen in Section~\ref{secdataexplanation}, the task of estimating the
parameters controlling the flux distribution and/or detecting a
breakpoint is additionally challenging because the fluxes depicted in
Figure~\ref{figsimplots} are not directly observed.

\subsection{Hierarchical modeling of the $\log N$--$\log S$ relationship}\label{secdataexplanation}
We now describe the connection between the broken power-law model
introduced in~(\ref{eqtwobreak}) and the observed data. In practice,
the flux of each source, $S_{i}$, is not observed directly. Instead, we
observe a Poisson-distributed photon count whose intensity is a known
function of the parameter $S_{i}$. Let $Y_1,Y_2,\ldots,Y_n$ denote the
source counts, then we assume the following hierarchical model. For
$i=1,\ldots,n$,
%
\begin{eqnarray}\label{eqnYmodel}
Y_i|S_1,\ldots,S_n & \stackrel{\mathrm{indep.}}{\sim}&\operatorname{Poisson}(A_iS_i +
b_i)\quad\mbox{and}
\nonumber\\[-8pt]\\[-8pt]\nonumber
S_i & \stackrel{\mathrm{i.i.d.}} {\sim}&\operatorname{Pareto}_B(\bolds{
\beta},\bolds{\tau }),
\end{eqnarray}
where $A_i$'s and $b_i$'s are known constants (see below), $\bolds{\beta
}=(\beta_1,\ldots,\beta_B)>\mathbf{0}$, $\bolds{\tau}=(\tau_1,\ldots,\tau_B)$ such that $\tau_B>\cdots>\tau_1>0$, and
$\operatorname{Pareto}_B(\bolds{\beta},\bolds{\tau})$ represents a $B$-piece Pareto
distribution with survival distribution
\[
S_B(x) = \cases{ 1, &\quad$x <\tau_1$,
\vspace*{3pt}\cr
\displaystyle
\biggl(\frac{\tau_1}{x} \biggr)^{\beta_1}, &\quad $\tau_1 \le x
< \tau_2$,
\vspace*{3pt}\cr
\displaystyle \biggl(\frac{\tau_1}{\tau_2}
\biggr)^{\beta_1} \biggl(\frac{\tau_2}{x} \biggr)^{\beta_2}, &\quad$
\tau_2 \le x < \tau _3$,
\vspace*{3pt}\cr
&\quad\vdots
\vspace*{3pt}\cr
\displaystyle \Biggl\{\prod^{B-1}_{j=1}
\biggl( \frac{\tau_j}{\tau
_{j+1}} \biggr)^{\beta_j} \Biggr\} \biggl(\frac{\tau_B}{x}
\biggr)^{\beta_B}, &\quad$x\ge\tau_{B}$,}
\]
and thus its distribution function $F_B(\cdot)=1-S_B(\cdot)$. Note
that the $B$-piece Pareto distribution corresponds to the broken power
law. The probability density $f_B$ can be easily found by
differentiation. When $B=1$, the $B$-Pareto distribution reduces to a
Pareto distribution with probability density function
\[
f_1(x; \beta,\tau) = \cases{ \displaystyle\frac{\beta\tau^\beta}{x^{\beta+1}}, &
\quad$x\ge \tau$,
\vspace*{5pt}\cr
0, &\quad$x<\tau$.}
\]
In the above $A_i$'s, sometimes known as effective areas, represent
sensitivities of the detector, while $b_i$'s represent background
intensities. With the above model the goal is then to estimate $B$ and,
at the same time, $\bolds{\beta}$ and $\bolds{\tau}$. At first sight,
this seems to be a straightforward statistical problem: for a fixed
$B$, maximum likelihood estimation can be used to estimate $\bolds{\beta
}$ and $\bolds{\tau}$, while the issue of choosing $B$ can be viewed as
a model selection problem and, thus, traditional ideas such as AIC and
BIC can be used. However, as to be seen below, practical implementation
of these ideas poses serious computational challenges that cannot be
easily solved.

\section{Maximum likelihood estimation when $B$ is known}\label{secestimation}

In this section we provide details of how to obtain maximum likelihood
estimates of $\bolds{\beta}$ and $\bolds{\tau}$ for a fixed number of
breakpoints $B$ in the $\log N$--$\log S$ model. Defining $\beta_0=0$,
$\tau_0=\tau_1$ and $\tau_{B+1}=\infty$, the likelihood is
\[
L(\bolds{\beta},\bolds{\tau};Y_1,\ldots,Y_n) = \prod
^n_{i=1} \biggl\{ \int^\infty_{\tau_1}
\frac{e^{-(A_is +b_i)}(A_i s+b_i)^{Y_i}}{Y_i!} f_B(s;\bolds{\beta},\bolds{\tau}) \,ds \biggr\}.
\]
Note that the likelihood involves some numerically unstable integrals
that do not have a closed-form solution and, hence, a direct
maximization is extremely difficult. To further appreciate this
difficulty, consider the case when there is no background contamination
($b_i=0$), for which the above likelihood degenerates to
\[
\prod^n_{i=1} \Biggl[\sum
^{B}_{j=1} \biggl(\frac{\tau_{j-1}}{\tau
_{j}}
\biggr)^{\beta_{j-1}}\frac{\beta_j(A_i\tau_j)^{\beta
_j}}{Y_i!} \bigl\{ \Gamma(Y_i-
\beta_j,A_i\tau_j)-\Gamma(Y_i-
\beta _j,A_i\tau_{j+1}) \bigr\} \Biggr].
\]
Here, $\Gamma(a,x) = \int_x^\infty t^{a-1} e^{-t}\,dt$ is the
incomplete gamma function which is numerically unstable, particularly
when the first argument is large. Together with the inner summation in
the above expression, these issues make a direct maximization of the
(log-)likelihood difficult even when there is no background
contamination. To address these issues, we propose an
EM-algorithm [\citet{DLR77}] to find the maximum likelihood
estimators of $\bolds{\beta}$ and $\bolds{\tau}$ for the general case of
$b_i\ge0$.

\subsection{EM with a sufficient augmentation scheme}
The EM algorithm\break [\citet{DLR77}] has long been popular for its
monotone convergence and resulting stability, and is therefore well
suited to our context. As always, the EM algorithm must be formulated
in terms of ``missing data'' or auxiliary variables, that must be
integrated out to obtain the observed data log-likelihood. For the
current problem, since we are interested only in inference for $\bolds
{\beta}$ and $\bolds{\tau}$,\vspace*{2pt} marginalizing over the uncertainty in the
individual fluxes, it is natural to treat $\mathbf{S} = (S_1,\ldots,S_n)^T$ as the missing data. Since $\mathbf{S}$ is a sufficient
statistic for $\bolds{\theta}=(\bolds{\beta},\bolds{\tau})^T$, we call
this the sufficient augmentation (SA) scheme in the terminology
of \citet{YuMeng11}.

Let $\mathbf{Y}=(Y_1,\ldots,Y_n)^T$. The complete data log-likelihood of
$(\mathbf{Y},\mathbf{S})$ is
\[
\log p(\mathbf{Y},\mathbf{S}; \bolds{\beta},\bolds{\tau})=\sum
^n_{i=1} \log g(Y_i;A_iS_i+b_i)
+ \sum^n_{i=1}\log f_B(S_i;
\bolds{\beta},\bolds{\tau}),
\]
where $g(x;\mu)$ is the probability mass function of a Poisson
distribution with mean~$\mu$.
In the E-step of the algorithm we compute the conditional expectation
%
\begin{eqnarray} \label{eqnQSAEM}
Q \bigl(\bolds{\theta}|\bolds{\theta}^{(k)} \bigr) &=& \mathbb{E} \bigl\{\log
p( \mathbf{Y},\mathbf{S};\bolds{\theta}) | \mathbf{Y}; \bolds{\theta }^{(k)}
\bigr\}\nonumber
\\
&=& \sum^n_{i=1} \mathbb{E} \bigl\{\log
g(Y_i;A_iS_i+b_i)|Y_i;
\bolds {\theta}^{(k)} \bigr\}
\\
&&{} + \sum^n_{i=1}\mathbb{E} \bigl\{\log f_B(S_i;\bolds{\theta})|Y_i;
\bolds{\theta}^{(k)} \bigr\},\nonumber
\end{eqnarray}
where $\bolds{\theta}^{(k)}$ denotes the estimate of $\bolds{\theta}$ at
the $k$th iteration. The M-step of the algorithm must then maximize
$Q(\bolds{\theta}|\bolds{\theta}^{(k)})$ with respect to $\bolds{\theta}$.
Since the first term of~(\ref{eqnQSAEM}) does not depend on $\bolds
{\theta}$, it can be ignored in our maximization. For the second term,
as it does not admit a closed-form expression, a Monte Carlo method is
used to approximate it. The basic idea is to estimate it by the mean of
a suitable Monte Carlo sample of the $S_i$'s as described in
Algorithm~\ref{algSAEM}.

\begin{algo}[t]
\caption{SAEM: EM with the sufficient augmentation scheme (SAEM)}\label{algSAEM}
\begin{longlist}[(3)]
\item[(1)] Choose a starting value $\bolds{\theta}^{(0)}$ and set $k=0$.

\item[(2)] Generate ${\mathbf{S}}^{(1)},\ldots,{\mathbf{S}}^{(N_{\mathrm{sim}})}$
from $p(\mathbf{S} | \mathbf{Y}; \bolds{\theta}^{(k)})$ using the following
Metropolis--Hastings algorithm. For each simulation of $\mathbf{S}$, we
sample the elements of $\mathbf{S}$ one at a time. Suppose $\mathbf
{S}=(S_1,\ldots,S_n)$ is the current draw. Denote $\mathbf
{S}^{*}=(S_1,\ldots, S_{j-1},S^{*}_{j},S_{j+1},\ldots, S_{n})$, where
$S^{*}_j$ is drawn from $\operatorname{Pareto}_B(\bolds{\beta}^{(k)},\bolds
{\tau}^{(k)})$. We accept this $\mathbf{S}^{*}$ as new value with
probability $a_j(\mathbf{S},\mathbf{S}^{*})$; otherwise, we retain~$\mathbf{S}$. The acceptance probability is given by
\begin{eqnarray*}
a_j \bigl(\mathbf{S},\mathbf{S}^{*} \bigr) &=& \min
\biggl\{ 1, \frac{g(Y_j;
A_jS^{*}_j+b_j)}{g(Y_j; A_jS_j+b_j)} \biggr\}.
\end{eqnarray*}
%

\item[(3)] Find the maximizer $\bolds{\tilde{\theta}}$ of the Monte Carlo
estimate of $Q(\bolds{\theta}|\bolds{\theta}^{(k)})$. This is equivalent
to computing
\[
\bolds{\tilde{\theta}} = \operatorname{argmax}\limits
_{\bolds{\theta}} \frac{1}{N_{\mathrm{sim}}-N_{\mathrm{burn}}}\sum
^{N_{\mathrm{sim}}}_{s=N_{\mathrm{burn}}+1}\sum^{n}_{i=1}
\log f_B \bigl(S^{(s)}_i;\bolds{\theta} \bigr),
\]
where $N_{\mathrm{burn}}$ is the number of burn-in. As discussed above,
$\tilde{\bolds{\theta}}$ can be obtained by the following steps:
\begin{enumerate}[(a)]
\item[(a)] set $\tilde{\tau}_1=\min\{S^{(s)}_i\dvtx  i=1,\ldots,n,s=N_{\mathrm{burn}}+1,\ldots,N_{\mathrm{sim}} \}$,
\item[(b)] obtain\vspace*{1pt} $\tilde{\tau}_2,\ldots,\tilde{\tau}_B$ as the
maximizer of $\sum^B_{j=1} m_j(\bolds{\tau}^{*})\log\beta_j(\bolds{\tau
}^{*})$, where $\bolds{\tau}^{*}=(\tilde{\tau}_1,\tau_2,\ldots,\tau
_B)$, using the Nelder--Mead algorithm, and
\item[(c)] set $\tilde{\beta}_j = \beta_j(\tilde{\bolds{\tau}})$
using~(\ref{eqnbetaj}), for $j=1,\ldots,B$.
\end{enumerate}
\item[(4)] Set $\bolds{\theta}^{(k+1)}=\tilde{\bolds{\theta}}$.
\item[(5)] Repeat steps (2) to (4) until convergence.
\end{longlist}
\end{algo}

Without the first term in~(\ref{eqnQSAEM}), the maximization of $Q(\bolds
{\theta}|\bolds{\theta}^{(k)})$ is equivalent to finding the MLE of
$\bolds{\theta}=(\bolds{\beta}, \bolds{\tau})^T$ from an i.i.d. sample $\mathbf
{X}=(X_1,\ldots,X_m)$ from the $\operatorname{Pareto}_B(\bolds{\beta},\bolds
{\tau})$ distribution. The log-likelihood of $\mathbf{X}$ is
\begin{eqnarray*}
l(\bolds{\theta};\mathbf{X}) & = & \sum^B_{j=1}
\beta_j ( n_j \log \tau_j -
n_{j+1} \log\tau_{j+1} ) + \sum^B_{j=1}
m_j \log \beta_j
\\
&&{} - \sum^B_{j=1} \beta_j
\sum_{i\in A_j}
\log X_i -
\sum^m_{i=1} \log X_i,
\end{eqnarray*}
where\vspace*{1pt} $n_j=\operatorname{card}\{i\dvtx  X_i\ge\tau_j\}$, $n_{B+1}=0$,
$m_j=n_{j+1}-n_j$, $\tau_{B+1}=\infty$,\break $n_{B+1}\log\tau_{B+1}$ is
defined to be 0, and $A_j=\{i\dvtx \tau_j\leq X_i<\tau_{j+1}\}$.
Note that the $n_j$'s and $m_j$'s are functions of $\bolds{\tau}$. For
any fixed $\bolds{\tau}$, straightforward\vadjust{\goodbreak} algebra shows that $l(\bolds
{\theta};\mathbf{X})$ is maximized when $\beta_j$ is set to
%
\begin{equation}
\beta_j(\bolds{\tau}) = {{m}_j(\bolds{\tau})} \biggl({\sum
_{i\in A_j
}\log X_i + {n}_{j+1}(\bolds{
\tau})\log{\tau}_{j+1}- {n}_j(\bolds{\tau }) \log{
\tau}_j} \biggr)^{-1}, \label{eqnbetaj}
\end{equation}
$j=1,\ldots,B$. By substituting the above expression, $l(\bolds{\theta
};\mathbf{X})$ becomes
%
\begin{equation}
l(\bolds{\theta};\mathbf{X}) = -m - \sum^m_{i=1}
\log X_i + \sum^B_{j=1}
m_j(\bolds{\tau}) \log{\beta}_j(\bolds{\tau}). \label{eqnloglikeX}
\end{equation}
Therefore, to obtain the MLE for $\bolds{\theta}=(\bolds{\beta},\bolds{\tau
})^T$ from $\mathbf{X}$, one can first maximize $l(\bolds{\theta};\mathbf{X})$
in~(\ref{eqnloglikeX}) with respect to $\bolds{\tau}$, and then plug
the corresponding maximizer $\hat{\bolds{\tau}}$ (i.e., the MLE of $\bolds
{\tau}$) into~(\ref{eqnbetaj}) to obtain the MLE $\hat{\bolds{\beta
}}$ for $\bolds{\beta}$.

The MLE of $\tau_1$ is $\hat{\tau}_1=\min(X_1,\ldots,X_m)$, while
unfortunately the MLEs for $\tau_2,\ldots,\tau_B$ do not admit
closed-form expressions. Further, (\ref{eqnloglikeX}) is not a
continuous function in $\bolds{\tau}$ and, therefore, traditional
optimization methods that require function derivatives (e.g.,
Newton-like methods) cannot be applied here. We have experimented with
various optimization algorithms and found that the Nelder--Mead
algorithm works well for this problem. The major steps of the EM
algorithm in the SA scheme (SAEM) for finding the MLEs of $\bolds{\theta
}$ are given in Algorithm~\ref{algSAEM}. In practice, the SAEM
algorithm often converges very slowly. Section~\ref{secemcompare}
below provides some illustrative numerical examples.

\subsection{EM with an ancillary augmentation scheme (AAEM)}
Given the slow convergence of the SAEM algorithm, we seek faster
alternatives. This subsection proposes an alternative EM algorithm that
is based on an ancillary augmentation (AA) scheme, called the AAEM
algorithm. For a discussion of augmentation schemes and their use in
EM, see~\citet{Baines12}. The basis of our AAEM is to re-express
our model using auxiliary variables $U_{i}=F_{B}(S_{i};\bolds{\theta})$:
\begin{eqnarray*}
Y_i|U_1,\ldots,U_n &\stackrel{\mathrm{indep.}} {\sim}& \operatorname{Poisson} \bigl(A_i
F_{B}^{-1}(U_i;\bolds{\theta})+b_i
\bigr) \quad\mbox{and}
\\
U_i &\stackrel{\mathrm{i.i.d.}} {\sim}& \operatorname{Uniform}(0,1),
\end{eqnarray*}
for $i=1,\ldots,n$. Here $\mathbf{U}=(U_1,\ldots,U_n)$ is treated as the
missing data and preserves the observed data log-likelihood. In the
E-step we then calculate the conditional expectation
%
\begin{equation}
Q \bigl(\bolds{\theta}|\bolds{\theta}^{(k)} \bigr) = \sum
^n_{i=1} \mathbb{E} \bigl\{\log g
\bigl(Y_i;A_iF_B^{-1}(U_i;
\bolds {\theta})+b_i \bigr)|Y_i;\bolds{\theta}^{(k)}
\bigr\}. \label{eqnQAAEM}
\end{equation}
This conditional expectation can be approximated and maximized in a
similar manner as for the $Q(\bolds{\theta}|\bolds{\theta}^{(k)})$ in the
SAEM algorithm. The resulting AAEM algorithm is summarized in
Algorithm~\ref{algAAEM}. Section~\ref{secemcompare} provides some
empirical comparisons between the AAEM and SAEM algorithms. As may be
expected, there are some situations where the AAEM algorithm converges
faster, while there are other situations where the SAEM algorithm
converges faster.

\begin{algo}[t]
\caption{AAEM: EM with ancillary augmentation scheme}\label{algAAEM}
\begin{longlist}[(1)]
\item[(1)] Choose a starting value $\bolds{\theta}^{(0)}$ and set $k=0$.

\item[(2)] Generate ${\mathbf{U}}^{(1)},\ldots,{\mathbf{U}}^{(N_{\mathrm{sim}})}$
from $p(\mathbf{U} | \mathbf{Y}; \bolds{\theta}^{(k)})$ using the
Metropolis--Hastings algorithm. For each simulation of $\mathbf{U}$, we
sample the element of $\mathbf{U}$ one by one. Let $\mathbf
{U}=(U_1,\ldots,U_n)$ be the previous draw. We denote $\mathbf{U}^{*}=(U_1,\ldots,
U_{j-1}, U^{*}_{j},U_{j+1},\ldots, U_{n})$, where $U^{*}_j$ is drawn
from $\operatorname{Uniform}(0,1)$. We accept this $\mathbf{U}^{*}$ as new
value with probability $b_j(\mathbf{U},\mathbf{U}^{*})$; otherwise, we
retain $\mathbf{U}$. The acceptance probability is given by
\begin{eqnarray*}
b_j \bigl(\mathbf{U},\mathbf{U}^{*} \bigr) &=& \min
\biggl\{ 1, \frac{g(Y_j;
A_jF^{-1}_B(U^{*}_j;\bolds{\theta^{(k)}})+b_j)}{g(Y_j; F^{-1}_B(U_j;\bolds
{\theta^{(k)}})+b_j)} \biggr\}.
\end{eqnarray*}

\item[(3)] Find the maximizer $\tilde{\bolds{\theta}}$ of the following
Monte Carlo estimate of $Q(\bolds{\theta}|\bolds{\theta}^{(k)})$:
\[
\frac{1}{N_{\mathrm{sim}}-N_{\mathrm{burn}}}\sum^{N_{\mathrm{sim}}}_{s=N_{\mathrm{burn}}+1}\sum
^{n}_{i=1}\log g \bigl(Y_i;A_iF_B^{-1}
\bigl(U_i^{(s)};\bolds{\theta} \bigr)+b_i \bigr).
\]
The maximization can be done, for example, with the Nelder--Mead algorithm.

\item[(4)] Set $\bolds{\theta}^{(k+1)}=\tilde{\bolds{\theta}}$.
\item[(5)] Repeat steps (2) to (4) until convergence.
\end{longlist}
\end{algo}

\subsection{Interwoven EM (IEM)}
In practice, choosing the most efficient algorithm between the SAEM and
AAEM requires knowledge of the unknown parameter values and the
theoretical convergence rates, both of which are not available in most
contexts. Therefore, it would instead be desirable if one could combine
the ``best parts'' of SAEM and AAEM rather than select one of them. One
simple way to combine the two algorithms is to use the so-called
alternating EM (AEM) algorithm. The AEM algorithm proceeds by using
SAEM for the first iteration, then uses AAEM for the second iteration,
followed by SAEM for the third, and so on. While this procedure tends
to ``average'' the performance of the two algorithms, a more
sophisticated way to combine them is to use the interwoven EM (IEM)
algorithm of~\citet{Baines12}. Theoretical and empirical results
show that IEM typically achieves sizeable performance gains over the
component EM algorithms. The key to the boosted performance of IEM is
that it utilizes the joint structure of the two augmentation schemes
through a special ``IE-step.'' In contrast, AEM simply performs
sequential updates using each augmentation scheme that makes no use of
this joint information. The theory of the IEM algorithm in~\citet
{Baines12} shows that the rate of convergence of IEM is dependent on
the ``correlation'' between the two component augmentation schemes.
Since the SA and AA schemes typically have low correlation, here we
interweave these two schemes to produce an IEM algorithm for estimating
the parameters of flux distributions.

\begin{algo}[t]
\caption{IEM: interwoven EM}\label{algIEM}
\begin{longlist}[(3)]
\item[(1)] Choose a starting value $\bolds{\theta}^{(0)}$ and set $k=0$.
\item[(2)] Execute steps~(2) and~(3) of the SAEM algorithm. Set $\bolds{\theta
}^{(k+0.5)}=\tilde{\bolds{\theta}}$.
\item[(3)] Execute step~(3) of the AAEM algorithm, with $\mathbf{U}^{(l)}$
generated as $U_j^{(l)}=F_B(S_j^{(l)};\bolds{\theta}^{(k+0.5)})$, for
$j=1,\ldots, n$ and $l=N_{\mathrm{burn}}+1,\ldots,N_{\mathrm{sim}}$. Set $\bolds
{\theta}^{(k+1)}=\tilde{\bolds{\theta}}$.
\item[(4)] If convergence is achieved or $k$ attains $N_{\mathrm{limit}}$, then
declare $\bolds{\theta}^{(k+1)}$ to be MLE; otherwise, set $k=k+1$ and
return to step (2).
\end{longlist}
\end{algo}

The IEM algorithm for our $\log N$--$\log S$~model is given in
Algorithm~\ref{algIEM}. The algorithm requires very minimal
computation in addition to the component SAEM and AAEM algorithms, so
is comparable in real-time per-iteration speed. Last, we note that
there is some freedom in how to combine the IEM algorithm with MC
methods. Specifically, there are variations in how one may choose to
implement step~(3). One may\vspace*{1pt} want to sample $U$ again instead of using the
previous samples in step~(2). In both cases, one obtains a sample from
$\mathbf{U}|\mathbf{Y},\bolds{\theta}^{(k+0.5)}$ and achieves the goal. From
our practical experience, we found that there is very little difference
between the performances of these two approaches. Thus, we choose to
use the one which is least computationally expensive.

\subsection{An empirical comparison among different EM algorithms}\label{secemcompare}

In this subsection we empirically compare the convergence speeds of
SAEM, AAEM, AEM and IEM by applying them to two simulated data sets.
These two data sets were simulated from a model with $B=1$ and no
background contamination counts. This model is somewhat simple, but the
advantage is that the likelihood function simplifies considerably, and
the corresponding maximum likelihood estimates can be reliably obtained
with non-EM methods. With these maximum likelihood estimates the
maximized log-likelihood value can be calculated and used for baseline
comparisons.

\begin{figure}

\includegraphics{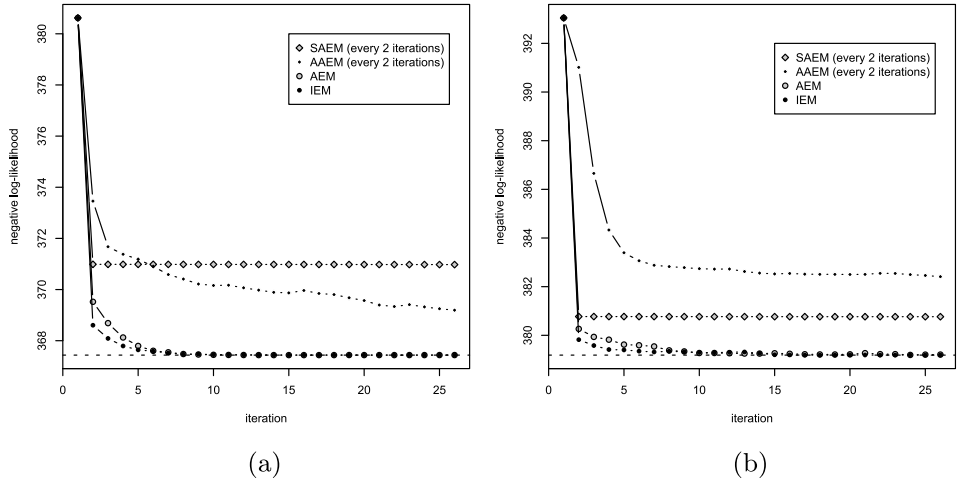}

\caption{Plots of negative log-likelihood values for different EM
algorithms. In each plot the horizontal dashed line indicates the
negative log-likelihood evaluated at the maximum likelihood
estimates.
\textup{(a)}~Simulated data set~1, \textup{(b)} simulated data set~2.}\label{figemcompare}
\end{figure}

In Figure~\ref{figemcompare}(a), for the first simulated data set, we
plot the negative log-likelihood values of the SAEM, AAEM, AEM and IEM
estimates evaluated at different iterations. One can see the slow
convergence speeds of SAEM and AAEM, with SAEM being the slower. Also,
both AEM and IEM converged relatively fast, with IEM being the faster.
When comparing to AEM, IEM utilizes the relationship between SAEM and
AAEM at each step, which leads to the superiority of IEM. As noted
earlier, the convergence rate of IEM is heavily influenced by the ``correlation'' between the two data augmentation schemes
being interwoven, that is, the SA and AA for this example. For the
$\log N$--$\log S$ model the correlation between these augmentation
schemes is hard to estimate exactly, but it appears empirically that
the SA and AA have a reasonably high correlation, thus preventing IEM
from outperforming AEM by a larger amount. This is likely due to $\tau
$, which controls the boundary of the parameter of the space and
heavily impacts the rate of convergence. However, among the candidate
algorithms IEM yields the best convergence properties.

%

We repeat the same plot in Figure~\ref{figemcompare}(b) for the second
simulated data set. This time the relative speeds of SAEM and AAEM
switched, that is, SAEM converged faster. This illustrates that neither
SAEM nor AAEM is uniformly superior to the other across all data sets.
The relative rate of convergence of AEM and IEM remain the same for
these two data sets and across other simulated data sets (not shown).

Overall, from these two plots one can see that the IEM algorithm is the
most efficient and robust. Also, when comparing to AEM, it is
computationally faster due to the skipping of an extra sampling step.
Similar performance was observed across a wide range of simulation
settings. Therefore, we recommend using the IEM algorithm to compute
the maximum likelihood estimates when $B$ is known.

\section{Automated choice of $B$}\label{secchoiceB}
This section addresses the important problem of selecting the number of
``pieces,'' $B$, in the broken-Pareto model. Since this problem can be
seen as a model selection problem, one can adopt well-studied methods
such as AIC and BIC to solve it. To proceed, we first note that when
$B=1$, the number of free parameters in the model is $2B$. With AIC,
the best $B$ is chosen as
\[
\hat{B}_{\mathrm{AIC}} = \operatorname{argmax}\limits
_{B} \mbox{AIC}(B) =
\operatorname{argmax}\limits
_{B} \bigl\{ -2 \log L(\hat{\bolds {\beta}},\hat{\bolds{
\tau}};Y_1,\ldots,Y_n) + 4B \bigr\},
\]
while for BIC $B$ is chosen as the minimizer of
\[
\hat{B}_{\mathrm{BIC}} = \operatorname{argmax}\limits
_{B} \mbox{BIC}(B) =
\operatorname{argmax}\limits
_{B} \bigl\{ -2 \log L(\hat{\bolds {\beta}},\hat{\bolds{
\tau}};Y_1,\ldots,Y_n) + 2B\log n \bigr\}.
\]
Despite the straightforward definitions, in practice, the numerical
instability of the likelihood function makes computation of $\mbox
{AIC}(B)$ and $\mbox{BIC}(B)$ very challenging. To address this
problem, we adopt the so-called power posterior method proposed by
\citet{Friel-Pettitt08} to approximate the log-likelihood directly.

In our context, the power posterior is defined as
\[
p_t(\mathbf{S}|\mathbf{Y};\bolds{\theta}) \propto p(\mathbf{Y} |
\mathbf{S})^t p(\mathbf{S};\bolds{\theta}) \qquad\mbox{for } 0\le t \le1.
\]
In addition, define
\[
z(\mathbf{Y}|t) = \int_{\mathbb{R}^n} p(\mathbf{Y}|
\mathbf{s})^t p(\mathbf{s};\bolds{\theta})\,d\mathbf{s}
\]
and, for simplicity, write the likelihood as $p(\mathbf{Y}) = L(\bolds
{\beta},\bolds{\tau};Y_1,\ldots,Y_n)$. The following equality is
crucial to this method:
\[
\log \bigl\{p(\mathbf{Y}) \bigr\} = \log \biggl\{ \frac{z(\mathbf
{Y}|t=1)}{z(\mathbf{Y}|t=0)} \biggr\} =
\int^1_0 \mathbb{E} \bigl[ \log \bigl\{p(
\mathbf{Y}|\mathbf{S}) \bigr\}| \mathbf{Y}; \bolds{\theta},t \bigr]\,dt,
\]
where the last expectation (inside the integral) is taken with respect
to the power posterior $p_t(\mathbf{S}|\mathbf{Y};\bolds{\theta})$.
The idea
is as follows. First, for any given $t$, Monte Carlo methods can be
applied to sample from the power posterior and approximate the
expectation. Once a sufficient number of these expectations
(corresponding to different values of $t$) are calculated, numerical
methods can be used to approximate the integral, which is the same as
the log-likelihood. Since this method approximates the log-likelihood
directly (i.e., without the computation of the likelihood), it is
numerically quite stable. The detailed algorithm is presented as
Algorithm~\ref{algpower}.

\begin{algo}[t]
\caption{Power posterior method for log-likelihood calculation}\label{algpower}
\begin{longlist}[(3)]
\item[(1)] Choose a starting value $\mathbf{S}^{(0)}$ and set $k=0$.
\item[(2)] Set $t=(k/N_{\mathrm{grid}})^c$, where $c$ controls the density of the
grid values of $t$. It is typically set to 3 or 5 [see \citet
{Friel-Pettitt08}].
\item[(3)] Generate ${\mathbf{S}}^{(1)},\ldots,{\mathbf{S}}^{(N_{\mathrm{sim}})}$
from $p_t(\mathbf{S} | \mathbf{Y}; \bolds{\theta})$ using the
Metropolis--Hastings algorithm described in step~(2) of the SAEM
algorithm. Note that the acceptance probability becomes
\begin{eqnarray*}
a_j \bigl(\mathbf{S},\mathbf{S}^{*} \bigr) &=& \min
\biggl\{ 1, \biggl\{\frac{g(Y_j;
A_jS^{*}_j+b_j)}{g(Y_j; A_jS_j+b_j)} \biggr\}^t \biggr\}.
\end{eqnarray*}
\item[(4)] Estimate $\mathbb{E} [ \log\{p(\mathbf{Y}|\mathbf{S})\}|
\mathbf{Y}; \bolds{\theta},t ]$ with
\[
\hat{l}_t=\frac{1}{N_{\mathrm{sim}}-N_{\mathrm{burn}}}\sum^{N_{\mathrm{sim}}}_{s=N_{\mathrm{burn}}+1}
\log p \bigl(\mathbf{Y}| \mathbf{S}^{(s)};\bolds {\theta} \bigr).
\]
\item[(5)] If $k< N_{\mathrm{grid}}$, set $k=k+1$, $\mathbf{S}^{(0)}=\sum^{N_{\mathrm{sim}}}_{s=N_{\mathrm{burn}}+1}\mathbf{S}^{(s)}/(N_{\mathrm{sim}}-N_{\mathrm{burn}})$, and go to step (2). Otherwise, go to the next step.
\item[(6)] Given the $\hat{l}_t$'s, the log-likelihood $\log\{p(\mathbf
{Y})\}
$ can be approximated via any reliable numerical integration method.
\end{longlist}
\end{algo}

The above algorithm provides a reliable method for approximating the
log-likelihood for a given value of $\bolds{\theta}$. Then one natural
question to ask is as follows: can we not simply obtain the MLE of $\bolds
{\theta}$ by directly maximizing this log-likelihood approximation
via, say, Newton's method? The answer, in principle, is yes, but the
IEM algorithm is still preferred mainly because the estimates from IEM
are generally more stable and reliable. Moreover, the power posterior
approximation to the log-likelihood is computationally intensive if one
wants to obtain an accurate estimate. For these reasons, we only use
this power posterior approximation to estimate the log-likelihood
evaluated at the MLE obtained by the IEM algorithm.

\section{Simulation experiments}\label{secSims}
Numerical experiments were conducted to evaluate the practical
performance of the proposed methodology. Four experimental settings
were considered:
\begin{longlist}[(1)]
\item[(1)] $B=1$, $\bolds{\tau}=5\times10^{-17}$, $\bolds{\beta
}=1$ and $n=100$,\vspace*{1pt}
\item[(2)]\label{S2} $B=2$, $\bolds{\tau}=(1\times10^{-17},5\times
10^{-17})^T$, $\bolds{\beta}=(0.5,3)^T$ and $n=200$,\vspace*{1pt}
\item[(3)]\label{S3} $B=2$, $\bolds{\tau}=(1\times10^{-17},5\times
10^{-17})^T$, $\bolds{\beta}=(0.5,1.5)^T$ and $n=200$,\vspace*{1pt}
\item[(4)] $B=3$, $\bolds{\tau}=(1\times10^{-17},8\times
10^{-17}, 1.8\times10^{-16})^T$, $\bolds{\beta}=(0.3,1,3)^T$ and $n=500$.
\end{longlist}
The parameter values of these settings were chosen to mimic the typical
behavior of the real data. The effective areas and the expected
background counts are set to $A_i=10^{19}$ and $b_i=10$, respectively,
for all $i$.

Two hundred data sets were generated for each experimental setting. For
each generated data set, both AIC and BIC were applied to choose the
value of $B$, and model parameters were estimated by the IEM algorithm.
The selected values of $B$ are summarized in Table~\ref{simB}. One can
see that BIC works substantially better than AIC for selecting $B$, and
while BIC occasionally overestimates $B$, there is a clear tendency for
AIC to consistently overestimate $B$.

Other crucial factors that determine the ability of our method to
detect structural breaks in the population distribution include: (i)
the sample size, (ii) the separation between breakpoints, and (iii) the
magnitude of the difference between the power-law slopes on adjacent
segments. The impact of the third factor can be seen by comparing
simulation results from settings~\hyperref[S2]{(2)} and~\hyperref[S3]{(3)}, where the
misclassification rate is seen to increase as the slopes become closer.
From additional simulations our experience suggests that in typical
settings a sample size of 200 or more is needed to reliably detect a
single breakpoint, with double this required to detect two breakpoints.
In simulations, true breakpoints can be detected for smaller sample
sizes, but at a lower rate that is more dependent on the noise
properties of the specific simulation.

%
\begin{table}[t]
\tabcolsep=8pt
\tablewidth=250pt
\caption{The number of pieces $\hat{B}$ selected by AIC and BIC}\label{simB}
\begin{tabular*}{\tablewidth}{@{\extracolsep{\fill}}@{}lcd{3.0}d{3.0}d{3.0}d{2.0}@{}}
\hline
\multirow{3}{52pt}{\\[-7pt] \textbf{Experimental setting}} & \multirow{3}{60pt}{\\[-14pt] \centering{\textbf{Model selection method}}} &\multicolumn{4}{c@{}}{$\bolds{\hat{B}}$}\\[-4pt]
& &\multicolumn{4}{c@{}}{\hrulefill}\\
& & \multicolumn{1}{c}{\textbf{1}} & \multicolumn{1}{c}{\textbf{2}} & \multicolumn{1}{c}{\textbf{3}} & \multicolumn{1}{c@{}}{\textbf{4}}\\
\hline
1&AIC & 94 & 53 & 35 & 18 \\
&BIC & 164 & 33 & 3 & 0\\[3pt]
2&AIC & 0 & 135 & 45 & 20\\
&BIC & 0 & 198 & 2 & 0\\[3pt]
3&AIC & 0 & 110 & 71 & 19\\
&BIC & 0 & 177 & 23 & 0\\[3pt]
4&AIC &0 & 0 & 138 & 62 \\
&BIC & 0 & 0 & 194 & 6 \\
\hline
\end{tabular*}
\end{table}

In addition to selecting the number of breakpoints, we also conducted a
simulation to assess the quality of parameter estimation when using the
IEM algorithm. For each experimental setting, we calculated the squared
error $(\beta_1-\hat{\beta}_1)^2$ of $\hat{\beta}_1$ for all those
data sets where $\hat{B}$ were correctly selected. We then computed
the average of all these squared errors, denoted as $\operatorname{m.s.e.} (\hat
{\beta}_1)$, and calculated the relative mean squared error $\sqrt
{\operatorname{m.s.e.}(\hat{\beta}_1)}/\beta_1$. Similar relative mean squared
errors for other estimates in $\hat{\bolds{\beta}}$ and $\hat{\bolds
{\tau}}$ were obtained in a similar manner. These relative mean
squared errors are given in Table~\ref{simest}. We note that all of
these are of the order of $10^{-2}$ or~$10^{-1}$.


\section{Application: \textit{Chandra} Deep Field North X-ray data}\label{secCDFN}

%
\begin{table}
\tabcolsep=0pt
\tablewidth=250pt
\caption{The relative mean squared errors of $\hat{\bolds{\beta}}$ and
$\hat{\bolds{\tau}}$, conditional on selection of the correct $B$. All
entries are multiplied by $10^2$}\label{simest}
\begin{tabular*}{\tablewidth}{@{\extracolsep{\fill}}@{}lc cd{2.2}cd{2.2}d{2.2}c@{}}
\hline
 &  &\multicolumn{3}{c}{$\bolds{\hat{\bolds{\tau}}}$} & \multicolumn{3}{c@{}}{$\bolds{\hat{\bolds{\beta}}}$}\\[-6pt]
& &\multicolumn{3}{c}{\hrulefill} & \multicolumn{3}{c@{}}{\hrulefill}\\
\textbf{Setting}& \textbf{Method}& \multicolumn{1}{c}{\textbf{1}} & \multicolumn{1}{c}{\textbf{2}} & \multicolumn{1}{c}{\textbf{3}} & \multicolumn{1}{c}{\textbf{1}} & \multicolumn{1}{c}{\textbf{2}} & \multicolumn{1}{c@{}}{\textbf{3}}\\
\hline
1&AIC &  5.14   & \multicolumn{1}{c}{--} & \multicolumn{1}{c}{--} &  11.1  & \multicolumn{1}{c}{--} &\multicolumn{1}{c@{}}{--} \\
&BIC &  4.91 &\multicolumn{1}{c}{--}&\multicolumn{1}{c}{--} &  10.6 &\multicolumn{1}{c}{--} &\multicolumn{1}{c@{}}{--}\\[3pt]
2&AIC &  3.33  &  2.55 &\multicolumn{1}{c}{--} &  9.81  &  11.3  &\multicolumn{1}{c@{}}{--}\\
&BIC &  3.52  &  2.60 &\multicolumn{1}{c}{--} &  9.17  &  10.8 &\multicolumn{1}{c@{}}{--} \\[3pt]
3&AIC &  3.52  &  14.2  & \multicolumn{1}{c}{--} &  12.0  &  13.2  & \multicolumn{1}{c@{}}{--}\\
&BIC &  3.57  &  12.9  & \multicolumn{1}{c}{--} &  11.1  &  13.5  & \multicolumn{1}{c@{}}{--}\\[3pt]
4&AIC &   2.71  &  3.26  &  5.04  &  7.08  &  9.91  &  12.3  \\
&BIC &  2.72  &  3.94  &  4.97  &  7.16  &  9.74  &  11.9  \\
\hline
\end{tabular*}
\end{table}

We now apply our method to data from the \textit{Chandra} Deep Field
North (CDFN) X-ray survey. Our data set comprises a total of 225
sources with an off-axis angle of 8 arcmins or less and counts ranging
from 5 to 8655. The full CDFN data set is comprised of multiple
observations at many different aimpoints, however, we here consider
only a subset where the aimpoints are close to each other to
avoid complications such as variations in detection probability due to
changes in the point spread function (PSF) shape and consequent
variations in detection probability.
The decision to include only aimpoints close to each other was taken
primarily to avoid the issue of ``incompleteness'' and
essentially amounts to taking a higher signal to noise subset of the
full data set. Incompleteness occurs when sources are not observed,
typically a result of being too faint to be detected under the specific
detector configuration used. Since this missingness is a function of
the quantity to be estimated, it must be accounted for and can lead to
tremendously more complicated and challenging modeling. This approach
is taken as part of a fully Bayesian analysis in~\citet
{BainesProc13}, but there are significant challenges to the method.
Most notably, results are very sensitive to the ``incompleteness function,'' which is frequently not known to such
high precision. By considering only a subset of aimpoints we focus on a
higher SNR subset of the \textit{Chandra} data that is not subject to
issues arising from incompleteness. We do not believe that the subset
choice impacts the final conclusion, as the results in the unpublished
report of Udaltsova, which models the full data set and accounts for
incompleteness, are extremely similar to those presented here.
Since the off-axis angle measures the radial distance of the source
from the center of the detector, sources with large off-axis angles can
be thought of as being ``close to the edge of the image.'' Sources
appearing at large off-axis angles appear much larger and at lower
resolution than those closer to the center of the detector. The
source-specific scaling constant, effective area $A_{i}$, is used to
account for variations in the expected number of photons as a function
of source location and photon energy. However, at large off-axis angles
additional complications such as ``confusion'' (two or more sources
overlapping and appearing as one) and ``incompleteness'' (possible
nondetections of fainter sources) must be considered.
For the purposes of our analysis here, we include all sources with an
off-axis angle $<$8~arcmin to achieve a worst-case completeness of $80\%
$. We also consider thresholding at $<$6 and $<$7~arcmins, with a full
discussion of the sensitivity to this threshold considered in
Section~\ref{subsecCDFNoffaxis}.

%
\begin{figure}[t]

\includegraphics{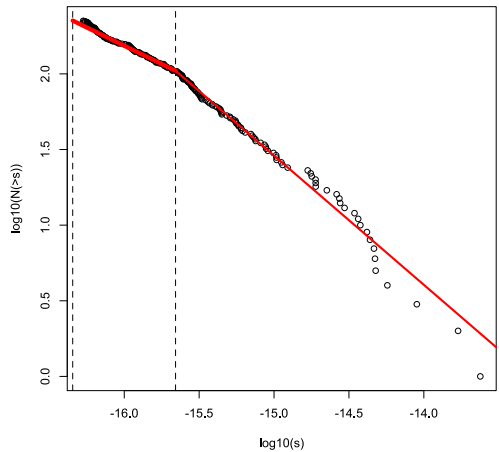}

\caption{$\log N$--$\log S$ plot for the \textit{Chandra} Deep Field
North data with off-axis angle truncation at 8~arcmins. The\vspace*{1pt} vertical
dotted lines are drawn at $\hat{\tau}_{1}$ and $\hat{\tau}_{2}$.
The red lines correspond to the fitted broken-Pareto model with
estimated slopes $\hat{\beta}_{1}$ and $\hat{\beta}_{2}$.}\label{figlnsplot}
\end{figure}

Applying our model selection procedure to the data set with
$<$8~arcmins yields an estimate of $\hat{B}=2$, with $\hat{B}=1$ for
the $<$6 and $<$7~arcmin subsets. As discussed in detail in
Section~\ref{subsecCDFNoffaxis}, the consistency of the observations
in the 6--8~arcmin range suggests that the ability to detect the
presence of a breakpoint is limited by the small sample sizes at $<$6
and $<$7. Figure~\ref{figlnsplot} shows the $\log N$--$\log S$~plot
for the $<$8~arcmin data set, depicting the log (base 10) of the
empirical survival count as a function of the log flux, using the
imputed fluxes from the final E-step of our algorithm. While the plot
ignores the uncertainty in the $S_{i}$'s, it remains the standard plot
for the analysis of $\log N$--$\log S$ relationships. We note from the
plot that the ``break'' is clearly visible around $\log_{10}(\tau
_{1})=-15.657$, with a change in slope from $0.48$ to $0.85$. Full
parameter estimates and standard error estimates are provided in
Table~\ref{tabcdfnresults}. Standard error estimates are obtained
using a simple Bootstrap resampling procedure. We also note that by
simulating from the model, the seemingly nonlinear behavior of the
curve at $\log(S)=-14.5$ is nonetheless seen to be consistent with the
piecewise linear model. Our analysis shows that a two-piece broken
power-law model is preferred for this subset, with a breakpoint at a
lower flux than shown in \citet{Moretti03} and with the lower
segment at a flatter slope. This differs from what would be expected if
point sources are to make up all of the diffuse background [\citet
{Hickox07}], suggesting that a significant proportion of the residual
X-ray background is composed of diffuse emission (e.g., hot
intergalactic plasma); see also \citet{Mateosetal2008}.

%
\begin{table}
\tabcolsep=0pt
\tablewidth=231pt
\caption{Parameter estimates and standard errors for the CDFN data set}\label{tabcdfnresults}
\begin{tabular*}{\tablewidth}{@{\extracolsep{\fill}}@{}ld{3.3}c@{}}
\hline
\textbf{Parameter} & \multicolumn{1}{c}{\textbf{Estimate}} & \textbf{SE} \\
\hline
$\beta_{1}$ & 0.483 & 0.060 \\
$\beta_{2}$ & 0.854 & 0.224 \\[3pt]
$\log_{10}(\tau_{1})$ & {-}16.344 & 0.030 \\
$\log_{10}(\tau_{2})$ & {-}15.657 & 0.271 \\
\hline
\end{tabular*}
\end{table}

The analysis in \citet{Hickox07} was based on optical sources
from the Hubble Space Telecope (HST) which had no X-ray counterparts.
By considering various models for the X-ray intensities of these
sources, \citet{Hickox07} compared them to the residual X-ray
background from deep Chandra observations. The proportion of the Cosmic
X-ray Background (CXB) that can be explained by point sources alone is
typically around 70--80\%. Connecting to our results, higher values for
$\beta_{1}$ increase the possibility that deeper observations could be
obtained that would explain an additional proportion of the CXB as
discrete sources. Alternatively, lower values for $\beta_{1}$ signify
a flatter $\log N$--$\log S$, suggesting a greater amount of diffuse
emission. Figure~8 of~\citet{Hickox07} depicts the relationship
between the proportion of the 0.5--2 keV CXB from unresolved HST point
sources and the power-law slope. The breakpoint estimated in our
analysis translates to $\approx 10^{-16}$~ergs$^{-1}$~cm$^{-2}$ for the passbands used by~\citet{Hickox07}. However,
in the 2 Msec data set they analyze, they do not detect any breakpoints
(see their Figure~7). Our analysis indicates that the $\log N$--$\log
S$ curve flattens for fluxes less than the breakpoint, thus allowing
for a significant proportion of the unresolved residual X-ray
background to be due to diffuse emission.

\subsection{CDFN source selection}\label{subsecCDFNoffaxis}
In this section we consider the sensitivity of our analysis to the
chosen off-axis angle threshold. As discussed in Section~\ref
{secCDFN}, at higher off-axis angles there are additional complications
such as incompleteness and confusion that must be built into any
statistical analysis that are not covered by the method presented here.
Let $K$ denote the maximum off-axis angle, that is, all sources with
off-axis angle less than $K$ are retained and all others are excluded
from the analysis. The choice of $K=8$ for our analysis in Section~\ref
{secCDFN} is motivated by scientific considerations and an estimated
completeness above 80\% at $K=8$. However, by varying the truncation
point we obtain additional insight into the sensitivity of our analysis
to this decision, as well as to the statistical sensitivity to the
sample size required for breakpoint detection. Table~\ref{tabCDFNK}
shows the results of the analysis for differing values of $K$. As
explained, results for $K>9$ are likely to be untrustworthy, although
they happen to be similar to those with $K=8$. On the other extreme, if
we truncate at $K=4$ or $K=5$, we unnecessarily discard a large number
of sources.

%
\begin{table}
\tabcolsep=0pt
\tablewidth=250pt
\caption{CDFN Results by varying off-axis truncation}\label{tabCDFNK}
\begin{tabular*}{\tablewidth}{@{\extracolsep{\fill}}ld{3.0}d{3.3}d{3.3}d{1.3}c@{}}
\hline
& & \multicolumn{2}{c}{$\bolds{\log_{10}(\hat{\tau})}$} & \multicolumn{2}{c@{}}{$\bolds{\hat{\beta}}$}\\[-6pt]
& & \multicolumn{2}{c}{\hrulefill} & \multicolumn{2}{c@{}}{\hrulefill}\\
$\bolds{K}$ & \multicolumn{1}{c}{$\bolds{n}$}& \multicolumn{1}{c}{$\bolds{\log_{10}(\hat{\tau}_{1})}$}&\multicolumn{1}{c}{$\bolds{\log_{10}(\hat{\tau}_{2})}$}
&\multicolumn{1}{c}{$\bolds{\hat{\beta}_{1}}$}&\multicolumn{1}{c@{}}{$\bolds{\hat{\beta}_{2}}$}\\
\hline
\phantom{0}4 &  77 & -16.364 &         & 0.788 & \\
\phantom{0}5 & 112 & -16.353 &         & 0.738 & \\
\phantom{0}6 & 152 & -16.329 &         & 0.691 & \\
\phantom{0}7 & 192 & -16.373 &         & 0.590 & \\
\phantom{0}8 & 225 & -16.343 & -15.668 & 0.482 & 0.850\\
\phantom{0}9 & 257 & -16.352 & -15.732 & 0.449 & 0.850\\
10 & 287 & -16.378 & -15.696 & 0.450 & 0.792\\
11 & 298 & -16.389 & -15.702 & 0.456 & 0.793\\
12 & 303 & -16.403 & -15.677 & 0.454 & 0.802\\
13 & 304 & -16.429 & -15.843 & 0.412 & 0.743\\
\hline
\end{tabular*}
\end{table}

We note that at $K=7$ we are also no longer able to formally detect a
break, that is, $\hat{B}=1$. However, upon closer examination the BIC
values for $B=1$ and  $B=2$ when $K=7$ are very similar (2186.79 vs.
2188.37), indicating that there is little to choose between the $B=1$
and $B=2$ models. With a few additional data points added at $K=8$, our
procedure then has enough power to detect the break at $K=8$. It is
worth noting that all additional data points with off-axis angle
between 7 and 8 were manually screened, and are quantitatively very
similar to those with $K<7$. That is, the detection (or lack) of a
breakpoint in this context appears to be primarily determined by the
sample size of the data set used. This is consistent with our results
from the simulation study in Section~\ref{secSims}, where a sample
size of approximately 200 was required to reliably detect a break with
similar parameter configurations. Indeed, looking at the plot in
Figure~\ref{figlnsplot}, we note that the break is rather a subtle
one, with the estimated slopes differing by approximately 0.37. In
summary, for this particular data set we note that there appears to be
evidence of a breakpoint, although the sample size required to detect
the breakpoint is not reached until we truncate at $K=8$, just before
additional modeling considerations such as incompleteness must be
accounted for.

%

\section{Theoretical properties}\label{seclargesample}

This section deals with the large-sample properties of the proposed
procedure. We first establish consistency results for the case when $B$
is known, with no background contamination ($b_i=0$ for all $i$) and
all $A_i$ are assumed to be identical. Then we describe how one could
weaken the assumptions of identical $A_i$'s and zero $b_i$'s. However,
as explained at the end of this section, the case of unknown $B$ is
substantially more difficult and we are unable to provide any
theoretical results for this case.

If it is assumed that $A_i=A>0$ and $b_i=0$ for all $i=1,\ldots,n$,
then $Y_1,\ldots, Y_n$ constitute an i.i.d. sample from model~(\ref{eqnYmodel}). Denote the density of $Y_1$ by
\begin{eqnarray*}
f(y; \bolds{\theta}) &=& \int^\infty_{\tau_1}
\frac
{e^{-As}(As)^{y}}{y!} f_B(s;\bolds{\beta},\bolds{\tau}) \,ds
\\
&=& \sum^{B}_{j=1} \biggl(
\frac{\tau_{j-1}}{\tau_{j}} \biggr)^{\beta
_{j-1}}\frac{\beta_j(A\tau_j)^{\beta_j}}{y!} \bigl\{ \Gamma (y-
\beta_j,A\tau_j)-\Gamma(y-\beta_j,A
\tau_{j+1}) \bigr\}. 
\end{eqnarray*}
The parameter space is defined as $\Theta= \{ \bolds{\theta} = (\bolds
{\beta},\bolds{\tau})^T\in\mathbb{R}^{2B}_{+}\dvtx \beta_j\neq
\beta_{j+1}, \tau_j<\tau_{j+1}, j=1,\ldots,B-1\}$. Let $\bolds{\theta
}_0=(\bolds{\beta}_0,\bolds{\tau}_0)^T\in\Theta$ denote the true
parameter value. Notice that $\Theta$ is not compact and that the
value of the likelihood does not converge to zero if the parameter
approaches the boundary of $\Theta$. Therefore, standard arguments
such as the ones based on \citet{Wald49} do not apply directly in
order to establish strong consistency of the maximum likelihood
estimator $\hat{\bolds{\theta}}$ of $\bolds{\theta}_0$. Instead a
compactification device is applied to subsequently use the results of
\citet{Kiefer-Wolfowitz56}. This leads to the following result.

%
\begin{teo}\label{teoconstA}
Suppose $B$ is known and $A_i=A>0$ for all $i=1,\ldots,n$. Then, the
maximum likelihood estimator $\hat{\bolds{\theta}}$ is strongly
consistent for $\bolds{\theta}_0$, that is, $\hat{\bolds{\theta}}\to\bolds
{\theta}_0$ with probability one as $n\to\infty$.
\end{teo}
The proof of Theorem~\ref{teoconstA} is provided in an online
supplement [\citet{Wong-Baines-Aue14}]. To weaken the restriction
of identical $A_i$, observe that this condition is mainly applied to
allow the use of the strong law of large numbers for i.i.d. random
variables, as required for the direct application of the results in
\citet{Wald49} and \citet{Kiefer-Wolfowitz56}. Since the
arguments used to prove Theorem~\ref{teoconstA} are still valid if
only the assumption $A_i>0$ is made, Kolmogorov's version of the strong
law of large numbers can be applied to adapt their proof to the present
case, imposing additional assumptions such as the Kolmogorov criterion
\[
\sum^\infty_{i=1}\frac{\operatorname{Var}(Y_i)}{i^2}<\infty
\]
or conditions ensuring the validity of Kolmogorov's three-series
theorem. Then, the result of Theorem~\ref{teoconstA} holds also in
this more general setting. The case for~nonzero $b_i$'s can also be
dealt with similarly, but with long and tedious algebra.

In the theory developed above, the number of pieces, $B$, in the
broken-Pareto model is assumed to be known. The case of unknown $B$ is,
however, substantially more difficult. In fact, in results in simpler
settings such as the traditional ``change in mean'' scenario, in which
segments of independent observations differ only by their levels,
strong distributional assumptions become necessary to show\vadjust{\goodbreak} consistency
of an estimator for $B$. These typically require normality of the
observations so that sharp tail estimates of the supremum of certain
Gaussian processes are available, for example, see \citet{Yao88}.
These techniques have also been exploited in \citet{Aue-Lee11}
for image segmentation purposes. However, in the current context of the
more complex broken-Pareto model, these arguments are not applicable
and, in fact, it seems infeasible to derive theoretical properties
under a set of practically relevant assumptions.

\section{Concluding remarks}\label{secConc}

We provide a coherent statistical procedure for selecting the number
and orientation of ``pieces'' in an assumed piecewise linear $\log
N$--$\log S$~relationship. Our framework allows astrophysicists to use
a principled approach to reliably select the model order $B$, and for
parameter estimation via maximum likelihood estimation in a numerically
challenging context. To our knowledge, this is the first statistically
rigorous procedure developed for solving this important scientific
problem. $R$ code implementing the proposed procedure can be obtained
from the authors.


\section*{Acknowledgments}
The authors are grateful to the Associate Editor and the Editor,
Professor Tilmann Gneiting, for their most useful and constructive
comments which substantially improved the paper.

\begin{supplement}
\stitle{Technical details}
\slink[doi]{10.1214/14-AOAS750SUPP} 
\sdatatype{.pdf}
\sfilename{aoas750\_supp.pdf}
\sdescription{We provide technical details of the proof of Theorem~\ref{teoconstA}.}
\end{supplement}



%

\printaddresses
\end{document}